\documentclass[aps,prb,article,groupedaddress,showpacs]{revtex4}

\usepackage{epsfig}
\usepackage{colordvi}
\usepackage{graphicx}
\usepackage{amssymb}
\begin{document}
\voffset 3cm
\title{
Universal behavior of internal friction in glasses below $T_g$ : anharmonicity vs relaxation}

\author{J. Pelous and C. Levelut}
\email{claire@lcvn.univ-montp2.fr}
\affiliation {Laboratoire des Collo\"{\i}des, Verres et Nanomat\'eriaux, CNRS/UMR5587,
Universit\'{e} Montpellier II, cc 69, 34095 Montpellier cedex, France}

\date{\today}
\begin{abstract}
 
Comparison of the internal friction  $Q^{-1}$  at hypersonic frequencies between a few K and
 the glass transition temperature  $T_g$  for various glasses brings out general features.
  At low temperature, $Q^{-1}$  is only weakly dependent on the material. At high
  temperature but still below $T_g$  the internal friction 
  for strong glasses shows a $T$-independent plateau in  a very wide domain of temperature; 
  in contrast,
   for fragile glass, a nearly linear variation of $Q^{-1}$  with $T$ is observed.
    Anharmonicity appears dominant over   
   thermally activated relaxational processes
    at high temperature.
\end{abstract}

\pacs{ 78.35.+c, 63.50+x, 63.20.Kr, 62.80.+f}
\maketitle

\section{Introduction}

Physical properties at low temperature, the mechanism of the glass transition 
 and the structure and vibrations at the nanometer scale account for many 
 experimental and theoretical contributions in 
 glasses.\cite{HunklingerArnold1976,Phillips1981,GotzeSjogren1992,FrickRichter1995,Angell2000,Elliot2001,PohlLiu2002,Angell1995}
 Search for universal behavior 
 or correlations between these properties are the mains fields explored. For 
 this purpose, it has been useful to classify glasses using the concept of fragility 
 \cite{Angell1991} depending on how the viscosity (or structural relaxation) 
 variation versus temperature deviates from Arrhenius behavior.
  One of the most debated questions 
 arises from the
  nature and the origin of the fast relaxation \cite{GotzeSjogren1992,FrickRichter1995} 
   responsible for a broad quasi elastic contribution 
  in neutron, X-ray and light 
  scattering and also for damping in sound waves and   dielectric loss.\cite{DyreOlsen2003}
  From analysis of 
   susceptibilities,
   a proportionality between quasi-elastic neutron 
   or light scattering, infrared absorption and  internal friction or sound 
   attenuation has been established and experimentally verified in a number of glasses. 
 \cite{TheodorakopoulosJackle1976,GilroyPhillips1981,BuchenauZhou1988,GurevichParshin1993,TerkiLevelut1997,FontanaRossi2005}

On the one hand, at very low temperature and for a large frequency range, many physical properties 
(among those  the internal friction $Q^{-1}$)
can be quantitatively described within the 
framework of the  ``tunneling model'' \cite{HunklingerArnold1976,Phillips1981} assuming a phenomenological 
potential with two asymmetric wells with a distribution of barriers and asymmetries. To improve
the description 
of data above  a few K the so called ``soft 
potential model''\cite{Parshin1994} 
extends and 
generalizes the previous  model 
 including distributed harmonic oscillators. 
At higher temperature, a phenomenological description of coupling of acoustic waves
 with unspecified thermally activated defects gives a  quantitatively correct description of
 the acoustical attenuation, or equivalently internal friction, at ultrasonic frequencies assuming 
distributions of energy barriers \cite{TielburgerMerz1992,DuquesneBellessa1985}.   
 Recently \cite{Buchenau2001}  this model has been 
  extended to describe  data for a large frequency range within the same formalism and to 
deduce the distribution of the energy barriers.  
Moreover some authors claim that inelastic light scattering results  can be
 accounted for by this description for many 
  order of magnitude in frequency in silica glass\cite{WiedersichAditchev2000} 
  as well as in various fragile
  glass.\cite{SurotsevWiedersich1988}
   In contrast, 
theoretical calculations explain acoustical damping in amorphous silicon and fused silica by
 anharmonicity for frequencies in the  10--100 GHz range.\cite{FabianAllen1999}. Furthermore, analysis 
 and discussion of data obtained using  picosecond 
 optical techniques in the same frequency range 
 conclude that
  classical relaxation theory cannot 
 explain the frequency and temperature variation observed.\cite{ZhuMaris1991} 
 Indeed, a description of the sound velocity (or elastic constants) versus temperature, even at low frequencies,
 cannot be accounted for assuming this relaxational process alone.\cite{Bellessa1972,ClaytorSladek1978,Nava1994,PaulGhosh1997}

On the other hand 
at hypersonic frequencies available from Brillouin scattering 
experiments, an attempt to describe experimental results for $Q^{-1}$ 
in silica glass  using the soft potential model \cite{BuchenauGalperin1992}  did
 not give satisfactory results when the temperature 
exceeds 10--20K. Moreover, comparisons of ultrasonic and hypersonic attenuation in glasses have
 demonstrated that thermally activated relaxations, dominant for temperatures higher than \mbox{10 K} at
 low frequencies,  cannot explain quantitatively the values observed 
at higher frequencies,
 not only in silica glass (Ref. ~\onlinecite{VacherPelous1981,Bonnet1991}) but also in  other
  glasses  (Ref. ~\onlinecite{BerretPelous1985,CutroniPelous1988}). 
Taking into account that the amplitude of sound  attenuation is not very different at high
 temperature in glasses and in crystals, anharmonicity has been invoked, by analogy with
 processes well known in crystals, to explain results at hypersonic frequencies. However, other authors
  \cite{TielburgerMerz1992} explain the same $Q^{-1}$ data in silica glass using the formalism
   of thermally activated processes.
   More recently careful analysis in silica glass have quantified the relative parts of 
   different processes responsible for
    the internal friction \cite{VacherCourtens2005} and demonstrate that anharmonicity 
    dominates at high 
    temperature.
Finally, the frequency dependence of the damping of vibrations 
 up to the THz range 
yield  conflicting 
interpretations.\cite{BenassiKrisch1996,RatForet1999,RuoccoSette1999,RuffleForet2003,RuzickaScopigno2004,RuffleGuimbretiere2006}
 Various frequency dependences
 at different temperatures are
 attributed to different origins such as relaxational, anharmonic or non-dynamic processes.
 Recently from measurements with
ultraviolet Brillouin light-scattering experiments  \cite{BenassiCaponi2005,MasciovecchioBaldi2006}
 questions about the structural or dynamic 
origin of sound scattering were again put forward.

So the description in the literature of the origin of the sound attenuation in glasses
 presents contradictory interpretations and it is not clear if the frequency and temperature
 variation of the internal friction can be interpreted in a formalism  common to all glasses. 
 Furthermore, most of the $Q^{-1}$  experimental
determinations concern temperatures lower than room temperature and most of the data 
below $T_g$ focus on 
fragile glass, and on silica glass as 
 representative of strong 
 glasses. As silica glass, like other tetrahedrally coordinated glasses, shows a number 
 of specific unexplained anomalies, 
 \cite{Vukcevich1972,Bruckner1970,HuangKieffer2005} 
 the question arises if the behavior observed 
 in silica is similar to that in other  glasses, as suggested in Ref.~\onlinecite{VacherCourtens2005}. 
To bring some light to these fields, we have made a comparison of results obtained
 for internal friction at hypersonic frequencies in a number of glasses. 
  This comparison reveals  new 
  characteristics common 
  to all materials at low temperature and in the temperature range below $T_g$.

\section{Experiments and results}

\begin{figure}[h]

\centerline{\epsfxsize=240pt{\epsffile{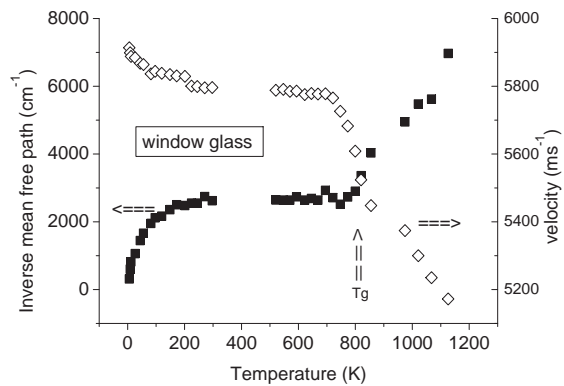}}}

\caption{\label{fig:fig1}Left axis: Longitudinal sound velocity versus temperature in a window  glass 
(see composition in Tab.~\ref{tab:tab1}) measured at hypersonic frequencies by Brillouin 
scattering in the
backscattering configuration with  
\mbox{$\lambda=514.5$ nm}. Right axis: inverse mean free path versus 
  temperature 
  in the same glass. 
}
 \end{figure}

The  general behavior of hypersonic properties of glasses is illustrated in 
Fig.~\ref{fig:fig1}  
 which gives results 
for the sound velocity 
and hypersonic attenuation, shown through the inverse mean free path, in window glass
 (72\% SiO$_2$, 14\% Na$_2$O, 
9\% CaO, 3\% MgO, 1\% Al$_2$O$_3$ plus other minor oxides).
 These data were obtained from Brillouin scattering experiments  with 
 experimental conditions 
 common to a series of previously published papers 
 \cite{VacherPelous1981,VacherSussner1980,VacherCourtens2005}.

   The Brillouin frequency  shift $\Delta \nu$ is related to the longitudinal sound wave velocity $V_l$, 
 the refractive index
  $n$ and the scattering angle $\theta$ by
  \begin{equation}
   \Delta \nu = 2n V_l(\sin{\theta/2})/ \lambda,
   \end{equation}
   where $\lambda$ is the wavelenght of the incident light.
   From the full width of the Brillouin line, $\Gamma$, one can deduce either the mean free path, $L^{-1}$,
     \begin{equation}
    L^{-1}= \Gamma \times 2 \pi / V_l,
    \end{equation}
      
or the internal friction, $Q^{-1}$, defined as 
\begin{equation}
 Q^{-1}=\Gamma/\Delta \nu.
 \end{equation}
 The accuracy of the experiments is about 0.1\% for the determination of sound velocity and 
 5--10\% for the mean free path (or the internal friction). \endnote {It 
  can also be
 noted that our results concerns only longitudinal waves for witch good accuracy can be obtained for the
Brillouin line-width determination but it has been pointed before \cite{VacherPelous1981} that the behavior is
	similar for transversal waves in this frequency range.}

\begin{figure}[h]

\centerline{\epsfxsize=210pt{\epsffile{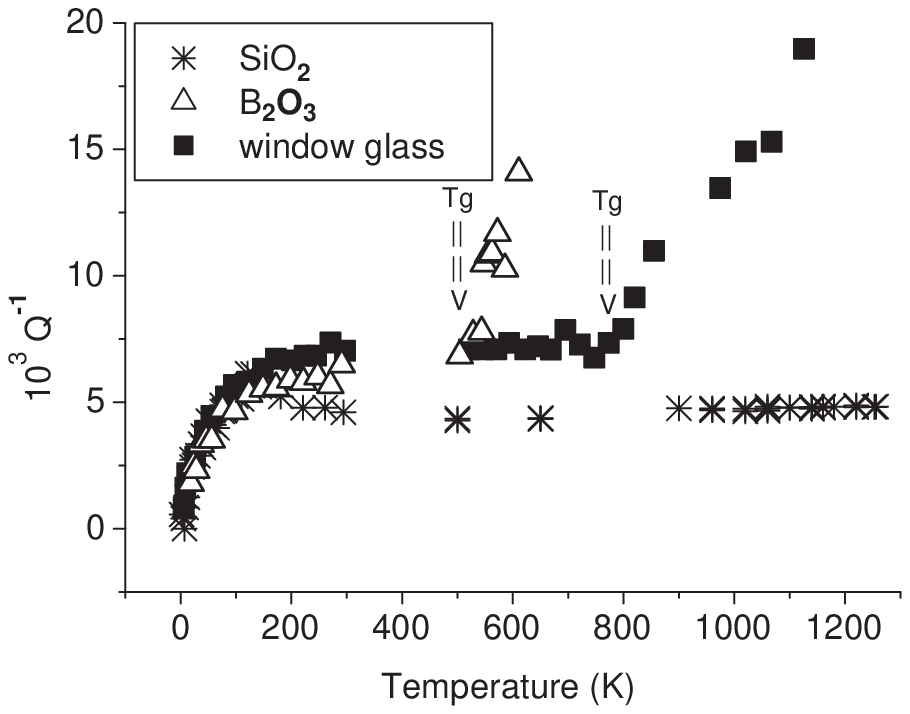}}}

\caption{\label{fig:fig2} Comparison of internal friction in strong glasses: silica 
(Ref.~\onlinecite{VacherSussner1980} and~\onlinecite{VacherCourtens2005} for the low temperature part and   Ref.~\onlinecite{LevelutLeParc2006} 
 for the high temperature part),  boron oxyde (Ref.~\onlinecite{Pelous1979} for the low temperature part and 
   Ref.~\onlinecite{LoroschCouzi1984} 
 for the high temperatures) and 
window glass.  }
\end{figure}

Fig.~\ref{fig:fig2} shows results in the full temperature range 
investigated for the two more standard 
glass formers, SiO$_2$  \cite{VacherSussner1980,VacherCourtens2005,LevelutLeParc2006}
and B$_2$O$_3$ \cite{Pelous1979,LoroschCouzi1984}
  together with 
  window glass.
  At least two characteristics common to these glasses 
 can be pointed out. First, in a large range of temperature,  $Q^{-1}$ is temperature independent 
  within the accuracy of the experiments. 
  Similar
  results have been found for other strong oxide glasses. \cite{TerkiLevelut1997,BerretPelous1985,VacherDelsanti1974}
    Secondly, 
  at the glass transition temperature,  a 
  strong increase of the internal friction is observed. This feature is associated with  the $\alpha$
   relaxational processes coupled to acoustic waves. A  signature at $T_g$ is also observed 
   for the sound velocity or other related elastic constants (Fig. \ref{fig:fig1}).

\begin{figure}[h]
\centerline{\epsfxsize=210pt{\epsffile{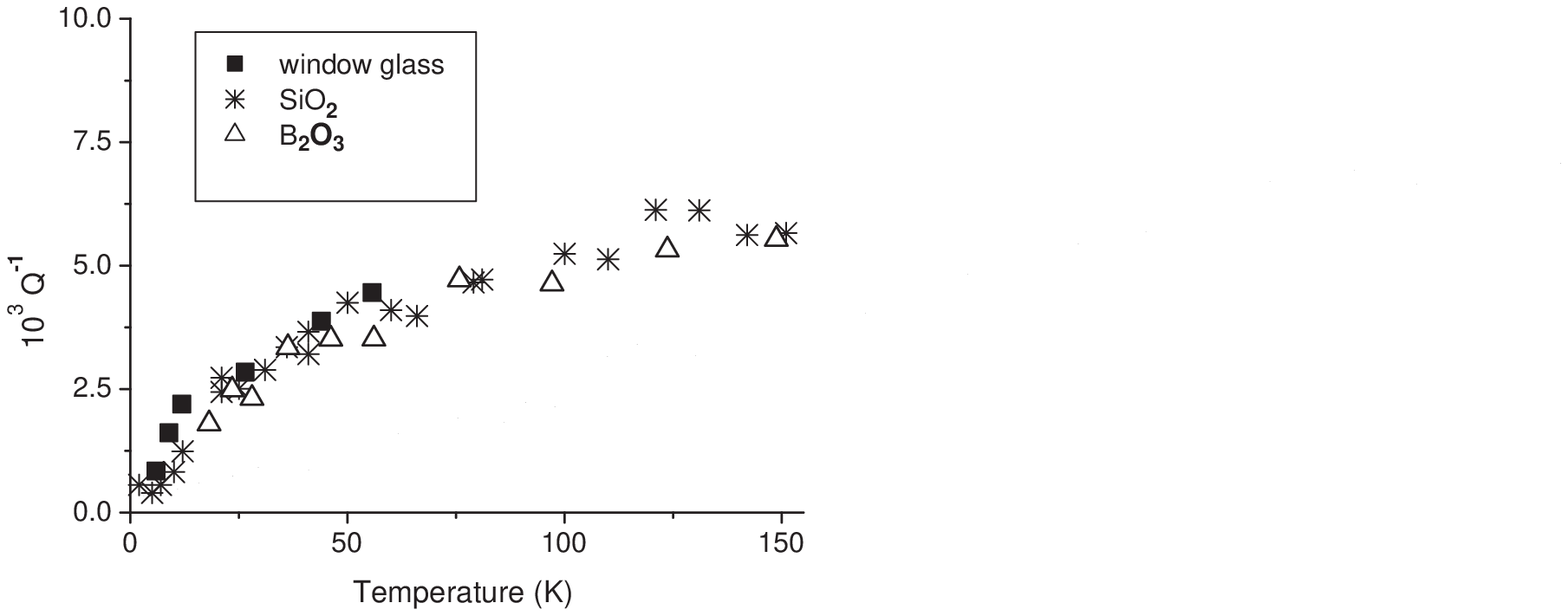}}}

\caption{\label{fig:fig3} Zoom on the low temperature part of internal friction in strong glasses as in Fig.
2.}
\end{figure}

\begin{figure}[h]
\centerline{\epsfxsize=240pt{\epsffile{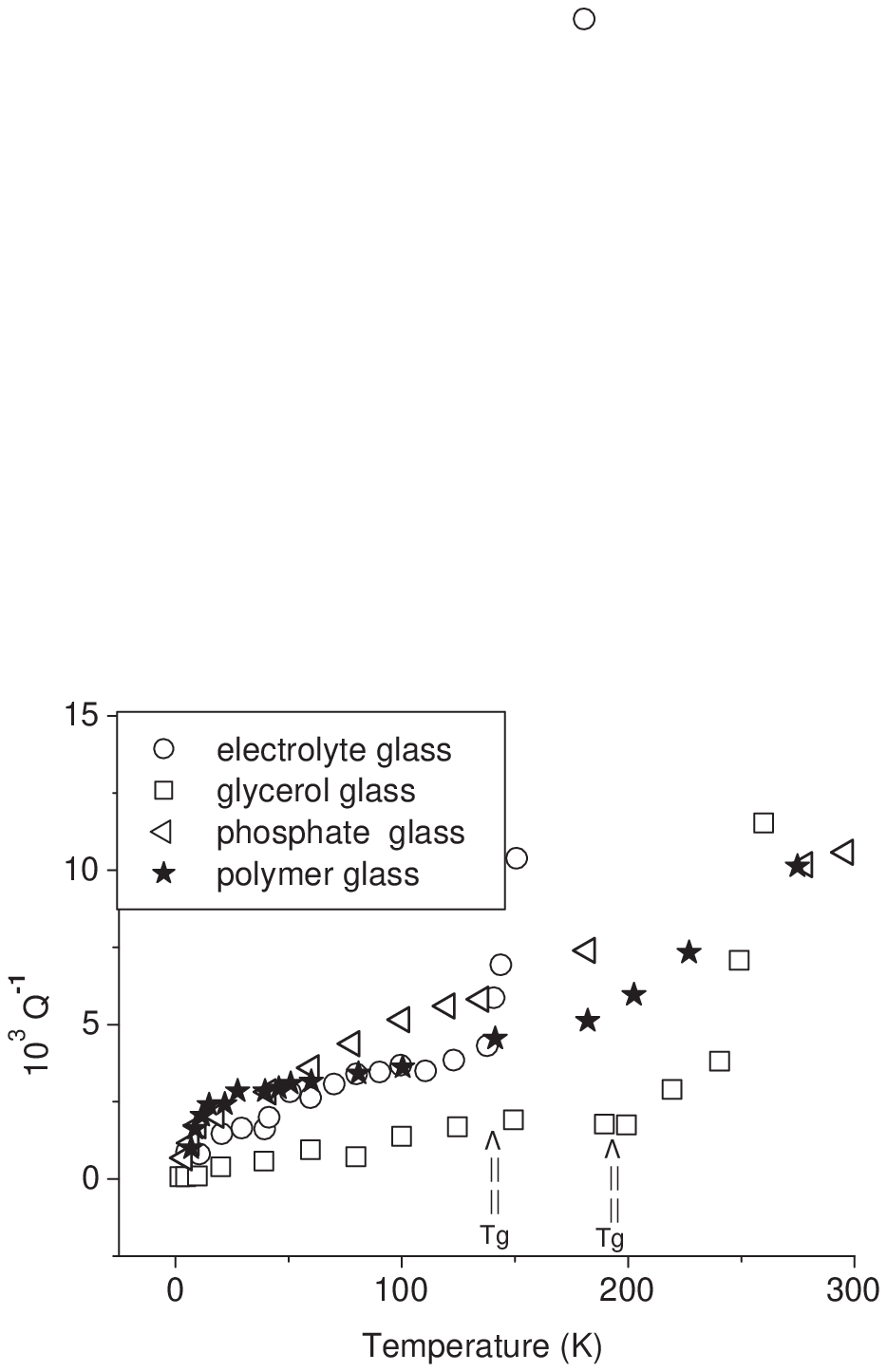}}}

\caption{\label{fig:fig4} Comparison of internal friction at low temperature in 
 an electrolytic glass LiCl-4H$_2$O \cite{PelousEssabouri1982}),  
 polymer PMMA (Ref. ~\onlinecite{VacherPelous1976}), phosphate glass (Ref.~\onlinecite{PelousVacher1980}) and
 glycerol (Ref. ~\onlinecite{VacherPelous1985}).
 }

\end{figure}

Another important feature  common to various glasses, not pointed out before, is displayed 
in    Fig.~\ref{fig:fig3} where results for temperatures lower than 150 K were considered. 
Below this temperature, the amplitudes 
of the internal friction tend to a common behavior  for oxide glasses as different as silica, boron oxide or window glass. 
 A peak reminiscent of the ultrasonic one, well identified in pure 
silica, does not appear for any other glass, including 
the fragile glasses shown in 
Fig.~\ref{fig:fig4}. In this  figure, $Q^{-1}$ values are compared for 
four examples of fragile glass: one electrolytic glass, LiCl-4H$_2$O, \cite{PelousEssabouri1982}
one organic polymer (PMMA)  \cite{VacherPelous1976}, glycerol \cite{VacherPelous1985}
and one   inorganic chain-like phosphate glass. \cite{PelousVacher1980} The amplitude of the internal friction is 
 similar in the strong window glass and in the fragile electrolyte glass, at low temperature, but significant differences between samples
 can be observed just below $T_g$.

\begin{table*}
\begin{tabular}{llccccccccc}	
&Chemical &$T_g$&$\Delta \nu$  &$ Q^{-1}$ &$\rho$&$V_d$&$V_l$&$C_v$&$\gamma$	 &calc. anharm. \\
& composition&(K)&(GHz)&$\times 10^3$&(g cm$^{-3}$)&(km s$^{-1}$)&(km s$^{-1}$)&(J cm$^{-3}$ K$^{-1}$)& &$10^3 \times Q^{-1}$\\
Silica glass	&		SiO$_2$	&		1400   &	33.6&		   4.8  &2.2&4.0&5.9&2.64&1.8 
&4.9 \footnote{The value of the anharmonic contribution in silica deduced by subtraction of the relaxational part
 is estimated to be equal to $Q^{-1}=
2.8  \times 10^{-3}$. \cite{VacherCourtens2005}}\\
Boron oxide glass   & B$_2$O$_3$			&   526	&20.0	& 		     6.8&1.8&2.0&3.5&2.16&2.7&	6.8\\

Polymer glass 	&polymethyl meta-	&	    370&	16.3	&		     10.1	&1.2&1.6&3.15&1.75&2.5&10.1\\ 
&crylate (PMMA)&&&&\\

\end{tabular}

		\caption{\label{tab:tab1} Comparison of experimental   $Q^{-1}$   and calculated anharmonic contribution of  $Q^{-1}$.	
	The experimental data are given at 300K. For silica and  boron oxide, the experimental values are identical at $T_g$ and room
	temperature. Data for SiO$_2$ are from
		 Ref.~\onlinecite{VacherSussner1980} and ~\onlinecite{LevelutLeParc2006},
		 data for B$_2$O$_3$ from 
		Ref.~\onlinecite{Pelous1979} and ~\onlinecite{LoroschCouzi1984}, data for PMMA from Ref.~\onlinecite{VacherPelous1976}.
		 Anharmonic contribution
		 $Q^{-1}$, calculated at $T_g$ for  SiO$_2$ and at 300K for  B$_2$O$_3$ and  PMMA. The calculated values
		 are
		  estimated from equation (\ref{eq:eq4}). The values of density, Debye and
		 longitudinal sound velocity,
		 specific heat and Gr\"uneisen parameter used for the claculation are also given in the table. We used 
		  $\tau_{ph}=10^{-13}$ s,  the Gr\"uneisen parameter
		   $\gamma$ from Ref.~\onlinecite{Novikov1998} and 
		  $C_v$ from Ref.~\onlinecite{BansalDoremus1986} for SiO$_2$ and PMMA, and 
		  from Ref.~\onlinecite{RamosMoreno1997}
		  for B$_2$O$_3$. }
		 \end{table*}

  \vspace{1cm}
 
 When the temperature increases, the internal friction 
continues to increase in fragile glasses whereas it levels off in strong glasses. This behavior has also been  
demonstrated in other polymeric glasses (polystyrene \cite{VacherPelous1981b}) 
or in an optical strong glass \cite{BerretPelous1985} above room temperature.

\section{Discussion}
As pointed in the introduction,   
 thermally activated relaxational processes have  been invoked to explain mechanical 
 damping in glasses below $T_g$. Those processes are useful to describe peaks observed at ultrasonic 
frequencies.  
As the same defects 
characteristic for the disorder are  partly responsible for the very low 
temperature properties,
 the 
distribution function of barriers and asymmetries of the tunneling model 
useful to describe
 internal 
friction and sound velocity below a few K are the starting point to 
consider thermally activated 
processes 
when the temperature is increased. \cite{HunklingerArnold1976,TielburgerMerz1992} Different hypothesis follow
 to extend the model  
and concern the 
form of the distribution functions. 
Assuming a flat distribution of asymmetries, $Q^{-1}$ is proportional to the imaginary 
part of the susceptibility,  and can be expressed as a function of 
   $g(V)$ 
    the temperature independent 
distribution function 
of barriers $V$
and $\tau=\tau_0\exp{V/kT}$ the relaxation time for hopping between adjacent 
potential wells \cite{TielburgerMerz1992,WiedersichAditchev2000}. 
 At hypersonic frequencies and  sufficiently high temperature
$\omega \tau_0 <<1 $  
 is satisfied,  $\tau_0$  being  the fastest relaxation time, and $\omega=2\pi\Delta \nu$.
Then 
\begin{equation}
Q^{-1} \simeq T  g( V). \label{eq:eq3}
\end{equation}  A maximum in $Q^{-1}$  in ultrasonic experiments implies a cut-off $V_{max}$ in $g(V)$. 
Different distributions $g(V)$ have been tested in the literature 
  but the 
general characteristic temperature dependence of $Q^{-1}$ can be discussed 
without a precise description  of $g(V)$.

The rough proportionality to  $T$ observed at hypersonic frequencies up to near \mbox{100 K} for strong glasses 
or below $T_g$ for other more fragile  glasses can be attributed to this process (Eq.\ref{eq:eq3}). The 
similarity of  values  obtained in different glasses can be related to the parameters of tunneling models deduced from 
 experiments at very low temperature which show close values
 for strong and fragile glass. \cite{BerretMeissner1988,PohlLiu2002} $Q^{-1}$ does not vary much for different glasses, about a
 factor of 2, as the  parameter characteristic for the contribution  of tunneling   defects to the internal friction, 
 as pointed out by Pohl.\cite{PohlLiu2002} 
 One can note the astoninshingly low value of  $Q^{-1}$ 
 for glycerol. 
 For most glasses, in ultrasonic experiments the $Q^{-1}$ peak is broader and appears 
 at higher temperature than in silica glass; \cite{HertlingBaebler1998}  a broadened distribution function with 
 higher potential barriers gives a good fit of the data but 
 calculations at
  hypersonic frequencies of the contribution of these thermally activated processes  using  the same 
  parameters 
   predict  an amplitude  much lower than experimental values
   \cite{VacherPelous1981,Bonnet1991,BerretPelous1985,VacherCourtens2005}
    or a maximum at a temperature higher than $T_g$,
  \cite{CutroniPelous1988} 
  so that another process, anharmonicity, 
   can be put forward at high temperature.

The second process under consideration is anharmonicity. 
It can be pointed out that the 
origin of this anharmonicity can be specific to glasses as developed by the authors of 
Ref~\onlinecite{Novikov1998,GurevichParshin2003,GotzeMayr2000} 
  but a quantitative test is not possible in the  currently available models. Furthermore, these authors  have pointed out that
  fragility is correlated with anharmonicity\cite{Novikov1998,GurevichParshin2003,GotzeMayr2000} and this contribution should occurs also in fragile
  glasses; this has been recently confirmed by molecular dynamic simulation. 
  \cite{BordatAffouard2004}
 
 The similarity between $Q^{-1}(T)$ curves in glasses and crystals at high temperature suggests that the same 
 formalism can be used in both materials.
 
In the  Akhieser regime \cite{Maris1971} using the same formalism 
of network viscosity as in crystals, the dominant process 
expected in glasses can be expressed as
\cite{FabianAllen1999}:
\begin{equation}
Q^{-1}= A \omega \tau_{ph}/ (1+ (\omega \tau_{ph}) ^2 ) \label{eq:eq4}
\end{equation}
with $A=\gamma^2  C_v  T  V_l / 2 \rho V_d^3$ where  
$C_v$ is the specific  heat per unit volume, $\gamma$ the Gr\"uneisen constant, $V_l$ the longitudinal sound velocity,
  $\rho$ the density and $V_d$ Debye velocity, $\tau_{ph}$ is the mean lifetime of thermal phonons.
More precise calculations of
  anharmonicity 
   \cite{FabianAllen1999} showed
that  a plateau is observed  at high frequencies. 

 At high temperature, but below $T_g$, $C_v$ in glasses presents
  a slow $T$ variation and the Debye model gives a good quantitative approximation; moreover in silicate glasses  $C_v$  value is 
  weakly dependent of the chemical composition.\cite{BansalDoremus1986} Assuming $\tau$ is proportional to $T^{-1}$
  (verified for example in various form of silica in 
  Ref. ~\onlinecite{VacherCourtens2005}) the product $C_v  T  \tau$  is a constant and can explain the leveling off for 
  the plateau $Q^{-1}(T)$ observed in oxide glasses.
Such a contribution also exist  in fragile glass and
  is superimposed on a relaxational one 
 which determine the observed  $T$ dependence. 
In order to quantify the relative part of these two contributions, the anharmonic contribution $Q^{-1}$ deduced from Eq. (\ref{eq:eq4})
 are calculated and 
given in  Tab.~\ref{tab:tab1} for the glasses for which the parameters are known or can be estimated. The Gr\"uneisen parameters
$\gamma$ are
taken from the article of Novikov, \cite{Novikov1998}  the specific heat per unit volume from Ref. ~\onlinecite{BansalDoremus1986} , 
Debye velocity ,$V_d$  is obtained from Pohl \cite{PohlLiu2002} or from our own measurements. $\tau_{ph}$ is  taken to be equal to
10$^{-13}$ s, as often in the literature.\cite{VacherCourtens2005} This lower limit for $\tau_{ph}$ provide a lower limit value for $Q^{-1}$.
Due to uncertainties on some parameters ($\gamma$ and $\tau_{ph}$) 
 only a rough estimate  
can be made but it appears that 
  an important part of the internal friction can be attributed to anharmonic process in all glasses from room
 temperature to $T_g$. Concerning the anomaly of glycerol, for temperatures lower than 150 K, the contribution of anharmonicity to
 internal friction is expected to be negligible (the condition $\omega \tau_{ph} << 1$ may not be fulfilled). 
 On the other hand,
the parameter characteristics for the tunneling amplitude  is the same as for silica,\cite{PohlLiu2002} so that we do not have an explanation for the 
 astonishingly low values of $Q^{-1}$ in glycerol.
 
Finally, to verify the consistency of the results, 
 the velocity variation vs temperature can be considered. This is not possible in silicate 
 glasses as structural anomalies dominate the elasticity.\cite{VacherCourtens2005}
  Below $T_g$  the variations $V(T)$
  in  glasses other than silica are linear. From the relative slope 
   $(1/V_l)(\delta V_l/\delta T)=\beta \gamma$ where $\beta$ is the thermal expansion,
    $\gamma$ is
   determined for the mode under study and can be different of $\gamma$ in Eq.~(\ref{eq:eq4}) 
   where $Q^{-1}$ is calculated
   in relation with the thermal bath of phonons. An 
  estimation of the Gr\"uneisen constant $\gamma$ of the mode can be deduced; we found $\gamma=3$ for both boron oxide
  and PMMA; these values  correlate well 
    with the  literature data used for the independent $Q^{-1}$ calculation (2.7 and 2.5, respectively from
   Ref.  ~\onlinecite{Novikov1998}).

\section{Conclusion}

Our contribution demonstrates that, at hypersonic frequencies accessible by
 Brillouin scattering experiments, the internal friction $Q^{-1}$ in strong oxide glasses is 
 basically $T$-independent
 above room temperature up to the glass transition. Moreover, the relaxation peak observed in 
 vitreous silica or  tetrahedrally
 coordinated glasses is not observed 
 in any other glass. \endnote{In GeO$_2$, analogue to SiO$_2$
  structure,
 $Q^{-1}$ shows also a very broad Brillouin peak near \mbox{300 K}; such a peak disappears in chemically
 multicomponent glasses.} 
 
 In strong and fragile glasses, $Q^{-1}$ differs from
  that of crystals mainly at low temperature and so the same mechanisms of sound 
  attenuation can be invoked 
 in these two classes of materials at high temperature. Additional contributions from
  activated relaxational processes specific 
 to glasses are efficient at low temperature and in more fragile glasses and can 
 explain the rough proportionality to $T$
 observed. The small variation with chemical composition of the internal friction 
 in the hypersonic regime below 100 K
 extend the remark about universality of low temperature properties  pointed out 
 in the review paper of Pohl toward high frequencies and toward higher temperatures than expected.\cite{PohlLiu2002}

\section*{Acknowledgments}The authors thanks
 D. Cavaill\'e  for 
 Brillouin scattering  experiments in window glasses and  Ian Campbell 
  for  fruitful comments  on the manuscript.

\bibliography{anharm.bib}
\bibliographystyle{apsrev}

\end{document}